\documentclass{PoS}

\usepackage{amsmath}
\usepackage{slashed}
\usepackage{wasysym}
\usepackage{graphicx}

\title{Detection prospects for conformally constrained vector-portal dark matter}

\ShortTitle{Detection prospects for conformally constrained vector-portal dark matter}

\author{\speaker{Frederick S. Sage}%
      \\
      University of Saskatchewan\\
      E-mail: \email{frederick.sage@usask.ca}}

\author{Zhi-Wei Wang\\
        $CP^3-Origins$, University of Southern Denmark and University of Saskatchewan\\
        E-mail: \email{wang@cp3.sdu.dk}}

\author{Rainer Dick\\
        University of Saskatchewan\\
        E-mail: \email{rainer.dick@usask.ca}}

\author{T.G. Steele\\
        University of Saskatchewan\\
        E-mail: \email{tom.steele@usask.ca}}

\author{R.B. Mann\\
        University of Waterloo\\
        E-mail: \email{rbmann@uwaterloo.ca}}

\abstract{We work with a UV conformal $U(1)^\prime$ extension of the Standard Model, motivated by the hierarchy problem and recent collider anomalies. This model admits fermionic vector portal WIMP dark matter charged under the $U(1)^\prime$ gauge group. The asymptotically safe boundary conditions can be used to fix the coupling parameters, which allows the observed thermal relic abundance to constrain the mass of the dark matter particle. This highly restricts the parameter space, allowing strong predictions to be made. The parameter space of several UV conformal $U(1)^\prime$ scenarios will be explored, and both bounds and possible signals from direct and indirect detection observation methods will be discussed.

Preprint Number: CP$^3$-Origins-2016-43 DNRF90}

\FullConference{38th International Conference on High Energy Physics\\
		  3-10 August 2016\\
		 Chicago, USA}

\begin{document}

\section{Extended $U(1)^\prime$ Vector-portal Dark Matter}

Observations suggest \cite{Bergstrom:2012fi} that the majority of matter in the universe is nonluminous and nonbaryionic, called dark matter. Extensions of the Standard Model (SM) of particle physics frequently include dark matter candidates in a dark sector that interacts with the SM through a `portal' of some kind. A well-established example is vector-portal dark matter, in which dark matter interactions are mediated by a vector boson, often associated with an extended gauge symmetry of some kind \cite{Leike:1998wr}\cite{Langacker:2008yv}. 

We work with a vector-portal model that includes an additional scalar $S$ charged under a new $U(1)^\prime$ gauge symmetry, which has a gauge field $B^\prime_\mu$ that gains a mass through radiative symmetry breaking via the Coleman-Weinberg mechanism \cite{Wang:2015sxe}\cite{PhysRevD.7.1888}, thus generating a $Z^\prime$ boson. Certain SM fermions $f$ have $U(1)^\prime$ charge and interactions also occur through kinetic mixing between $B^\prime_\mu$ and the SM hypercharge field. By including a Higgs-portal interaction between the new scalar and the SM Higgs, the radiative symmetry breaking in the hidden sector (singlet sector) triggers electroweak symmetry breaking in the Higgs sector. The model is furnished with an additional Dirac fermion $\chi$ charged under $U(1)^\prime$ that can act as cold dark matter. Additional vector-like spectator fermions $\psi$ and a set of right handed neutrinos $\nu_R$ are required for anomaly cancellation \cite{Carena:2004xs}\cite{PhysRevD.91.035025}, but do not contribute to the phenomenology. To avoid stringent dilepton constraints \cite{Aaboud:2016cth} on $U(1)^\prime$ models, we work with a leptophobic variant \cite{PhysRevD.91.035025}. The relevant extension of the Lagrangian is reported in \cite{Wang:2015sxe}. To constrain the model, we fix the mass of the $Z^\prime$ boson to $m_{Z^\prime}=1.9$ TeV, in agreement with a possible LHC resonance \cite{Aad:2015owa}. This constraint will be relaxed in a future article.

\section{UV Boundary and Thermal Constraints}

By using the renormalization group, we can constrain the gauge and scalar sector couplings to reduce the parameter space of the model. To realize asymptotic safety in the scalar sector, we require that the SM Higgs quartic coupling (and possibly other scalar couplings) reach a fixed point at some UV scale $\Lambda_{UV}$, which provides certain UV boundary conditions and  generates a stable Higgs vacuum. The renormalization group equations are solved with these boundary conditions, and the resulting couplings at the electroweak scale are presented in \cite{Wang:2015sxe}. We discuss in particular two scenarios, A and B, which have respectively $U(1)^\prime$ gauge couplings 0.18 and 0.1, and mixed gauge couplings 0.034 and 0.045.

Assuming dark matter is thermally produced as in the WIMP paradigm, its abundance is governed by the rate equation, which takes a simplified form under the analytic Lee-Weinberg approximation \cite{weinberg1977}. Solving the rate equation and inserting the observed dark matter abundance \cite{Ade:2015xua} creates a constraint on the thermally averaged annihilation cross section.

The constrained cross section can be compared against the cross section calculated using the gauge coupling values computed from the UV boundary conditions to restrict the mass range of the model to a handful of points. This comparison appears in Figure 1. The masses where the curves intersect the abundance constraint are the only mass values that are acceptable as cold dark matter. Our model is highly predictive, admitting thermal dark matter only at a set of masses near $m_{Z^\prime}/2$ (Scenario A - 855 GeV, 1004 GeV; Scenario B - 880 GeV, 980 GeV).

\begin{figure}
\centering
\includegraphics[scale=0.3]{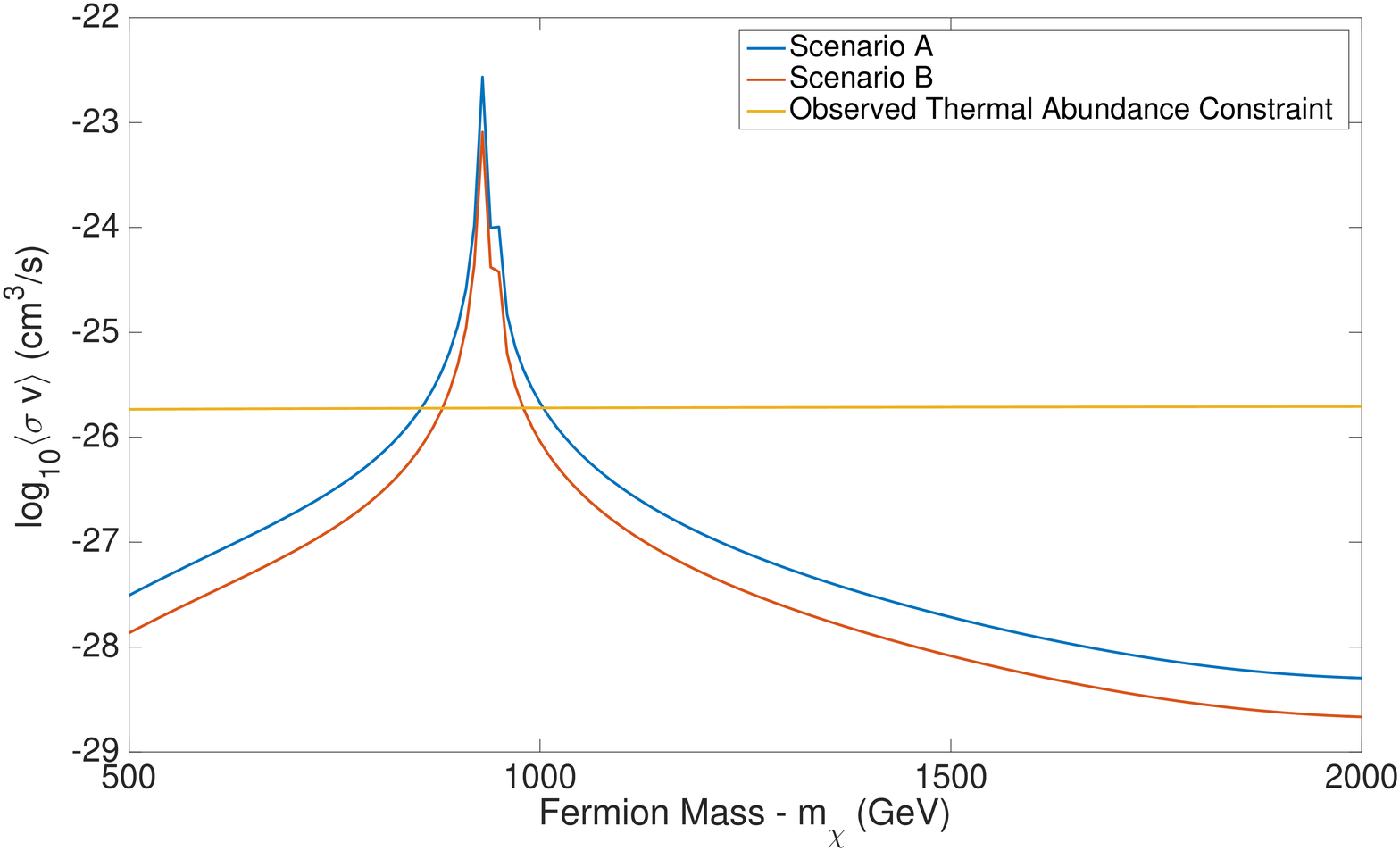}
\caption{Thermal constraint on annihilation cross section}
\end{figure}

\section{Direct and Indirect Detection Limits}

The nuclear recoil cross section for vector-portal dark matter is readily available in the literature \cite{Cline:2014dwa}. The strongest experimental constraints on the properties of heavy (>100 GeV) dark matter that exist at the current time are from the direct nuclear recoil searches. The recoil cross sections of our models are compared against current bounds by LUX \cite{Faham:2014hza}. These exclusion bounds rule out the thermally admitted masses for Scenario A.

Indirect searches for dark matter are those that look for products of dark matter annihilation in the galactic halo. One of the most promising signals is the gamma ray signal, due to the fact that photons are easy to detect and travel in straight lines, allowing for spatially targeted searches. The photon spectra that result from the annihilation of dark matter are found in \cite{Cirelli:2010xx}.

Dark matter in the Galactic Halo is expected to contribute to the Isotropic Gamma Ray Background through annihilations. We compare the expected flux from annihilation in our model using standard halo input parameters to the observed flux by the Fermi space telescope \cite{Ackermann:2014usa} using the ratio of the predicted flux over the observed flux on a logarithmic scale. Results are presented for the two scenarios in Figure 2. For a ratio larger than unity, the predicted flux is larger than what is observed, ruling out the mass value.

Our results indicate that the higher mass value for Scenario A is inconsistent with the Fermi observations and that the higher mass value for Scenario B is potentially observable.

\begin{figure}
\centering
\includegraphics[scale=0.3]{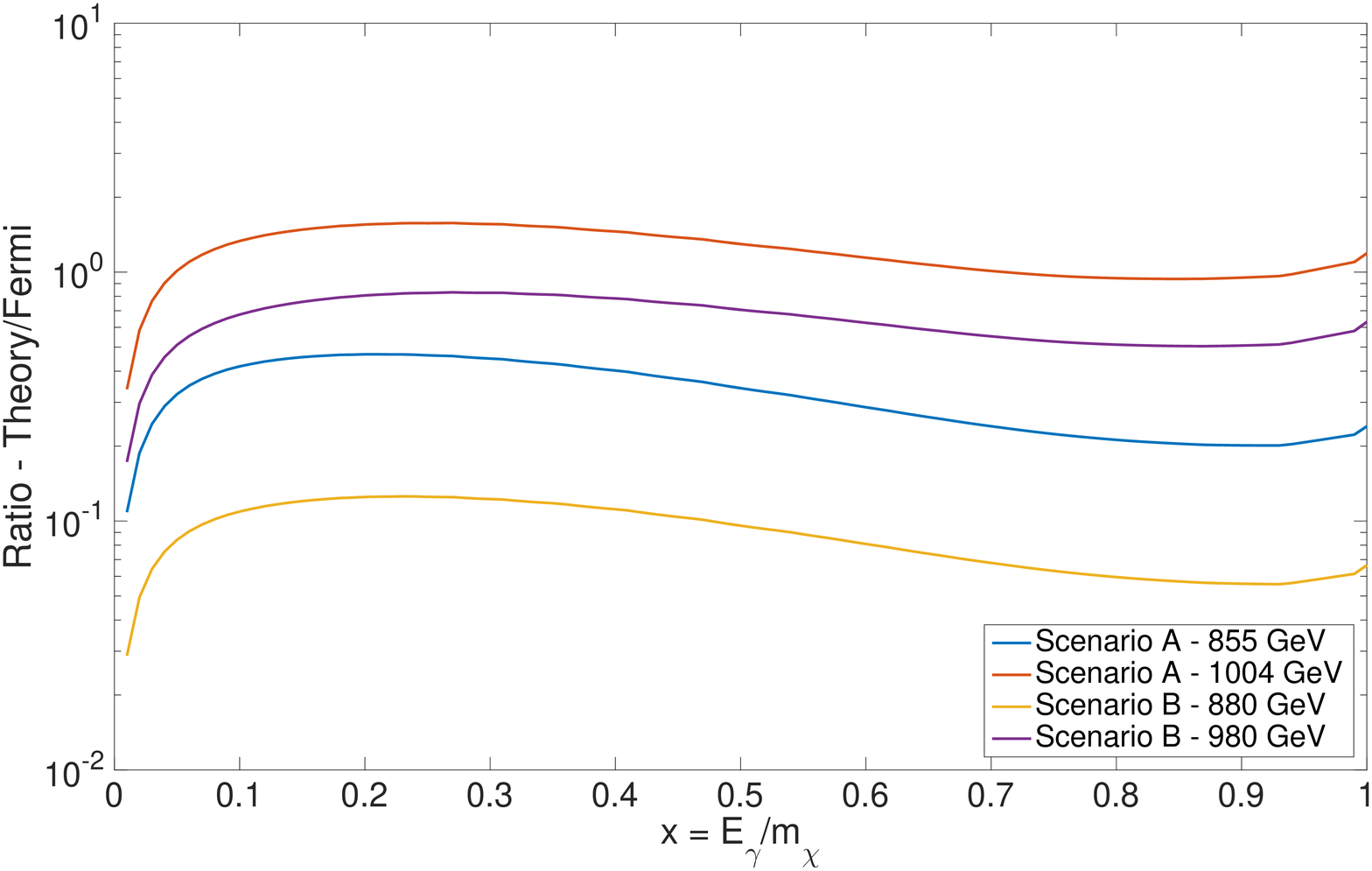}
\caption{Ratio of predicted flux over observed flux}
\end{figure}

\section{Conclusions}

We have explored the detection prospects of thermal relic vector-portal dark matter in two different UV boundary scenarios. The UV boundary conditions fix the coupling parameters and the thermal constraint fixes the mass to a handful of points. Further comparison with direct and indirect exclusion bounds rules out one of the scenarios. The other remains potentially viable.

{\bf Acknowledgments}- This work was supported in part by the NSERC Discovery Grant Program.

\bibliographystyle{h-physrev}
\bibliography{Ichep}

\end{document}